\documentstyle[epsfig]{jaa}
\begin{document}
\font\jnk=cmr10 scaled\magstep2 
\title[Mauritius Radio Telescope]{A low frequency radio telescope at Mauritius for a Southern sky
survey}
\author[Golap, Udaya Shankar, Sachdev, Dodson, Sastry]{K. Golap$^1$, N. Udaya Shankar$^{2,1}$, S. Sachdev$^{2,1}$,
\and R. Dodson$^{2,4}$, Ch. V. Sastry$^{3,1}$ \\
$^1$Department of Physics, University of Mauritius, Reduit, Mauritius\\
$^2$Raman Research Institute, Sadashivanagar, Bangalore 560080, India\\
$^3$Indian Institute Of Astrophysics, Koramangala, Bangalore 560034, India\\
$^4$Physics Department, University of Durham, Durham, U.K.} 
\maketitle
   
\begin{abstract}
           A new, meter-wave radio telescope has been built in the
North-East of Mauritius, an island in the Indian ocean, at a latitude
of $-20.14^{\circ}$. The Mauritius Radio Telescope (MRT) is a Fourier
Synthesis T-shaped array, consisting of a 2048~m long East-West arm and
a 880~m long South arm. In the East-West arm 1024 fixed helices are
arranged in 32 groups and in the South arm 16 trolleys, with four
helices on each, which  move on a rail are used. A 512 channel
digital complex correlation receiver is used to measure the visibility
function. At least 60 days of observing are required for obtaining
the visibilities up to 880~m spacing. The Fourier transform of the
calibrated visibilities produces a map of the area of the sky
under observation with a synthesized beam width $4' \times 4.6'$
sec$(\delta+20.14^{\circ})$ at 151.5~MHz.

The primary objective of the telescope is to produce a sky survey in
the declination range $-70^{\circ}$ to $-10^{\circ}$ with a point
source sensitivity of about 200~mJy ($3 \sigma$ level). This will be
the southern sky equivalent of the Cambridge 6C survey. In this paper
we describe the telescope, discuss the array design and the
calibration techniques used, and present a map made using the
telescope.
\end{abstract}        

\begin{keywords} 
Radio Telescope -- Low Frequency -- Imaging -- Southern Sky
-- Fourier Synthesis
\end{keywords}


\newpage   

PLATE I. An Aerial view of the Mauritius Radio Telescope. The observatory building is also seen. \\

PLATE II.  A closer view of the helices used.
\newpage  
     
\section{Introduction}      
Surveying the sky and compiling catalogs of celestial objects has been
a major part of astronomical research for centuries. The first
systematic survey of the radio universe was carried out by Grote Reber
\cite{reber} using a backyard telescope with a resolution of
12$^{\circ}$ operating at a frequency of 160~MHz. With the quest for
higher angular resolution the exercise of surveying soon shifted to
higher frequencies. Even so, many low frequency surveys were carried
out after  Reber's survey some of which are summarized in
Table~\ref{tab:tab_1}. The table clearly indicates that the sixth
Cambridge survey (6C) \cite{6C_paper1} is by far the most extensive
survey at low frequencies. This survey provides a moderately deep
radio catalog reaching a source density of about $2 \times 10^4~{\rm \
sr}^{-1}$ over most of the sky north of $\delta=+30^{\circ}$ with an
angular resolution of $4.2' \times 4.2'$ cosec$(\delta)$ and a
limiting flux density of 120~mJy at 151~MHz. An equivalent of the 6C
survey for the southern sky does not exist.

Since the survey of Mills {\it et al.} \cite{Mills} at 80~MHz there
has not been much effort at low frequencies to map the southern sky
apart from the Parkes 408~MHz survey. The Culgoora \cite{culgoora}
observations at 80 and 160~MHz were made mainly to study known
sources and determine their spectral indices. The largest existing radio survey
of the southern sky is the Parkes-MIT-NRAO survey \cite{PMN} at
5~GHz. There is an obvious need to survey the southern sky at a
frequency around 150~MHz. At this frequency synchrotron sources show
up much better than at higher frequencies, such as 408~MHz, due to
their spectra. Sources also show up better at 150~MHz than at lower
frequencies since the absorption due to the interstellar gas is much
less than at deca-meter wavelengths. For this
purpose a radio telescope operating at 150~MHz has been constructed at
Bras d'Eau, Mauritius.

\begin{table}[b]
\footnotesize
\begin{tabular}{|c|c|c|c|c|c|}
\hline \multicolumn{6}{|c|}{Northern declination surveys} \\ \hline
\hline \multicolumn{1}{|c|}{Freq.}  &\multicolumn{1}{|c|}{Observatory}
&\multicolumn{1}{|c|}{Resol.}  &\multicolumn{1}{|c|}{Dec. Coverage}
&\multicolumn{1}{|c|}{Sensit.}  &\multicolumn{1}{|c|}{No. of Sources}
\\ \hline 
408~MHz & Effelsberg & $0.85^{\circ} \times 0.85^{\circ}$ & $ -10^{\circ}
 +50^{\circ}$ & 0.2~Jy & --\\ 
\hline 
178~MHz & Cambr. 3CR & $2' \times 2'$ &
$ -5^{\circ} +90^{\circ}$ & 9~Jy & $328$ \\ 
\hline 
178~MHz & Cambr. 4C & $ 7.5'
\times 7.5'$ & $-7^{\circ} +80^{\circ} $& 2~Jy & $4843$\\ 
\hline 
38~MHz & Cambr. WKB & $ 45' \times 45'$ & $-45^{\circ} +35^{\circ} $ & 14~Jy &
$1000$ \\
\hline 
34.5~MHz & GEE TEE & $ 30 ' \times 30' $ & $ -30^{\circ} +60^{\circ} $ & 5
Jy & $\approx 3000$\\ 
\hline 
151~MHz & Cambr. 6C & $ 4' \times 4' $ & $ +30^{\circ}
+90^{\circ} $ & 0.12~Jy & $>10^5$\\ 
\hline
\multicolumn{6}{|c|}{Southern declination surveys}
\\ \hline \hline
38~MHz & Cambr. WKB & $ 45' \times 45'$ & $-45^{\circ} +35^{\circ} $ & 14~Jy & $1000$ 
\\ \hline
34.5~MHz & GEE TEE & $ 30 ' \times 30' $ & $ -30^{\circ} +60^{\circ} $ & 5~Jy &
$\approx 3000$ \\
\hline
408~MHz & Parkes & $0.85^{\circ} \times 0.85^{\circ}$ & $ -60^{\circ} +10^{\circ}$ & 1~Jy & --\\
\hline
408~MHz & Molongolo & $2' \times 2'$ & $ -60^{\circ} +18^{\circ}$ & 0.6~Jy & $>12000$\\
\hline
843~MHz & MOST & $0.4' \times 0.4'$ & $-90^{\circ} -30^{\circ}$ & -- & ongoing\\
\hline
151.5~MHz & MRT & $4' \times 4'$ & $-70^{\circ} -10^{\circ}$ & 0.2~Jy & ongoing
\\
\hline
\hline
\end{tabular}\\

\caption{Surveys below 1 GHz}
\label{tab:tab_1}
\end{table}

\begin{figure}[h] \epsfig{file=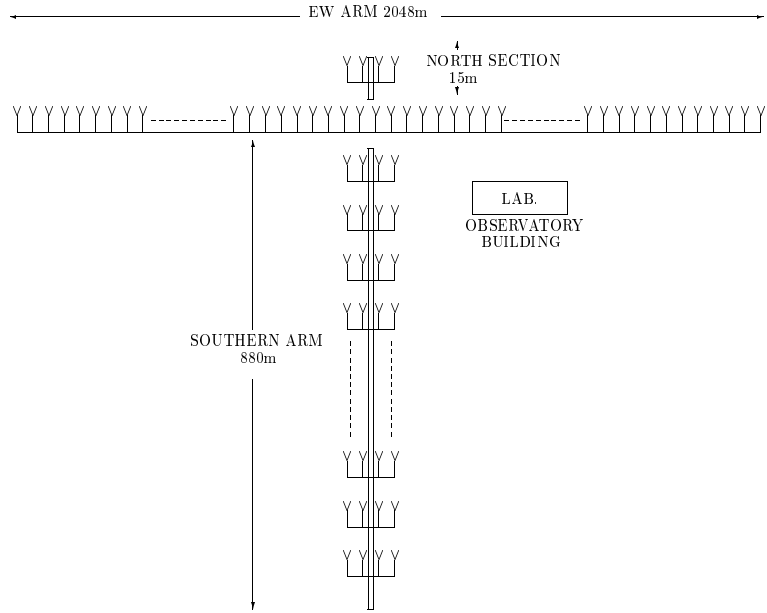,height=10cm,width=12.5cm}
\caption{{\footnotesize A schematic of the array. The EW arm has 1024
helices which are divided in 32 groups of 32 helices each. The NS arm
has 16 trolleys each with 4 helices.}} \label{fig:fig_1} \end{figure}

The Mauritius Radio-Telescope (MRT) has been constructed and is
operated collaboratively by the Raman Research Institute, the Indian
Institute of Astrophysics and the University of Mauritius. It is
situated in the North-East of Mauritius (Latitude 20.14$^{\circ}$
South, Longitude 57.74$^{\circ}$ East), an island in the Indian
Ocean. It is a T-shaped array with an East-West (EW) arm of length
2048~m having 1024 helical antennas and a South (S) arm of length
880~m consisting of a rail line on which 16 movable trolleys each with
four helical antennas are placed (Figure~\ref{fig:fig_1}).
The helical antennas respond to frequencies between 80 and
160~MHz. Presently the telescope is operated at 151.5~MHz which allows
maximum interference-free observations. The helices are mounted with a
tilt of $20^{\circ}$ towards the south so that they point towards a
declination of $-40^{\circ}$ at the meridian. The declination coverage
corresponding to the Half Power Beam Width (HPBW) of the helices is
from $-70^{\circ}$ to $-10^{\circ}$. Plate I shows an aerial view of
the telescope and Plate II shows a closer view of the helices used.

The EW arm is divided into 32 groups of 32 helices each. All the EW
groups are not at the same height, a situation imposed by the uneven
terrain. Each trolley in the S arm constitutes one S group. Both EW
and S group outputs are heterodyned to an intermediate frequency (IF)
of 30~MHz, using a local oscillator (LO) at 121.6~MHz. The 48 group
outputs are then amplified and brought separately to the observatory
building via coaxial cables. In the observatory, the 48 group outputs
are further amplified and down converted to a second IF of
10.1~MHz. The 32 EW and 16 S group outputs are fed into a $32 \times
16$ complex, 2-bit 3-level digital correlator sampling at 12~MHz. The
512 complex visibilities are integrated and recorded at intervals of 1
second. At the end of 24 hours of observation the trolleys are moved
to a different position and new visibilities are recorded. A minimum
of 60 days of observing are needed to obtain the visibilities up to
the 880~m spacing.

The Fourier transform of the phase corrected visibilities obtained
after the complete observing schedule produces a map of the area of
the sky under observation with a synthesized beam width of $4' \times
4.6'$ $\sec(\delta + 20.14^{\circ})$. The phase corrections mentioned
above take into consideration the non-coplanarity of the
baselines. The expected root mean squared (RMS) values of the
background in the synthesized images, arising from the system noise
with a 1~MHz bandwidth and an integration time of 8 seconds and from
the confusion noise are expected to be around 200~mJy(3$\sigma$) and
10~mJy respectively.

Table~\ref{tab:mrt_specs} summarizes the MRT specifications.

\begin{table}
\begin{center}
\begin{tabular}{l l}
\multicolumn{1}{l} {Observing frequency} &\multicolumn{1}{l}{151.5 MHz}\\

\multicolumn{1}{l} {Telescope configuration} &\multicolumn{1}{l}{T shaped 2048 m EW arm and 880 m NS arm}\\
\multicolumn{1}{l} {Basic element} &\multicolumn{1}{l}{helical antenna}\\
\multicolumn{1}{l} {Polarization} &\multicolumn{1}{l}{Right Circular}\\
\multicolumn{1}{l} {HPBW of helix} &\multicolumn{1}{l}{$60^{\circ}\times60^{\circ}$}\\
\multicolumn{1}{l} {Declination coverage} &\multicolumn{1}{l}{-70$^{\circ}$ to -10$^{\circ}$}\\
\multicolumn{1}{l} {Collecting area of helix} &\multicolumn{1}{l}{4 m$^2$ at 150 MHz}\\
\multicolumn{1}{l} {East-West arm} &\multicolumn{1}{l}{32 groups with 32 helices each}\\
\multicolumn{1}{l} {North-South arm} &\multicolumn{1}{l}{15 trolleys each with 4 helices}\\
\multicolumn{1}{l} {$1^{st}$ IF frequency} &\multicolumn{1}{l}{30 MHz}\\
\multicolumn{1}{l} {$2^{nd}$ IF frequency} &\multicolumn{1}{l}{10.1 MHz}\\
\multicolumn{1}{l} {Instrumental bandwidths} &\multicolumn{1}{l}{0.15,1.0,1.5,3.0 MHz}\\
\multicolumn{1}{l} {Digitization before correlation} &\multicolumn{1}{l}{2-bit 3-level}\\
\multicolumn{1}{l} {Correlation receiver} &\multicolumn{1}{l}{32 $\times$ 16 complex channels}\\
\multicolumn{1}{l} {No of baselines measured per day} &\multicolumn{1}{l}{$32 \times 16$}\\
\multicolumn{1}{l} {Minimum and maximum baselines} &\multicolumn{1}{l}{0, 1024 $\lambda$}\\
\multicolumn{1}{l} {Time to get full resolution image} &\multicolumn{1}{l}{60 days}\\
\multicolumn{1}{l} {Synthesized beam-width} &\multicolumn{1}{l}{$4' \times 4.6'$sec$(\delta+20.14^{\circ})$}\\
\multicolumn{1}{l} {Point source sensitivity} &\multicolumn{1}{l}{200 mJy (3 $\sigma$)}\\
\end{tabular}
\caption{MRT Specifications}
\label{tab:mrt_specs}
\end{center}
\end{table}

\section{The Telescope}

\subsection{Design criteria}
\label{sub_sec-design_crit}
The array was designed by considering, the availability of the 512
channel Clark Lake correlator system\footnote{ After the closure of
the Clark lake radio-telescope the two-bit three-level 512 channel
complex correlation receiver was kindly donated to the
MRT. }\cite{CLRO}, the constraints due to the available terrain and
the presence of man-made interference.

\paragraph{Configuration} A T-shaped configuration was chosen for its
simplicity. Aperture synthesis with fixed antennas in the EW arm and
movable elements (trolleys) in the S arm were chosen to minimize the
hardware required. An abandoned old railway line running North-South
was rebuilt for use as the South arm. This new rail line, slopes
downwards at about $ 1/2 ^{\circ}$ to the horizontal till about 655 m,
and then slopes upwards at about $1 ^{\circ}$ to the horizontal. On
this rail the trolleys cannot approach the EW arm nearer than 11 m.

To ensure that the array responds to structures of all sizes in the
sky, the array should provide all spacings available in a square
aperture. To meet this requirement at MRT, a 15~m North extension with
one trolley almost touching the EW arm has been built and is used to
measure the low spatial frequencies.  However, this trolley cannot
approach the EW array nearer than 2~m.  Hence baselines with 1~m
spacing in the S direction are not measured.  Non-zero EW baselines
with zero-spacing along the S~direction are obtained by multiplying
the groups of the eastern arm with the output of four helices which
are a part of the first group (closest to the center) of the western
arm. This has the same primary beam as a trolley and ensures a
weighting similar to baselines with non-zero spacing along the S
direction.

\begin{figure}[h]

\epsfig{file=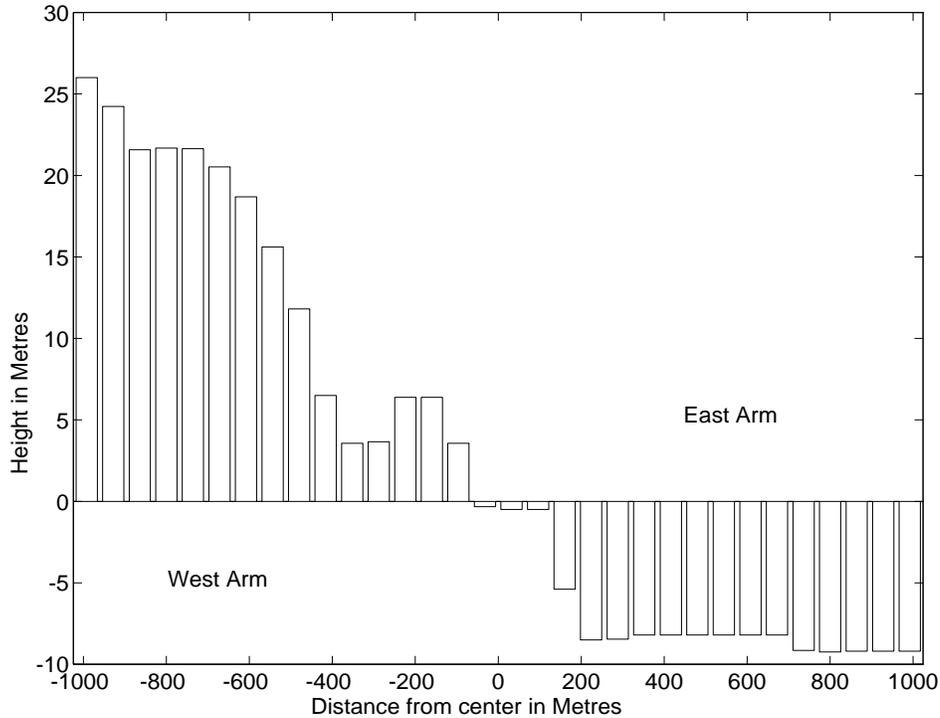,height=10cm,width=12.5cm}
\caption{{\footnotesize The EW height profile. The maximum height difference in the
EW arm is 35 m.}}
\label{fig:fig_2}
\end{figure}

\paragraph{Terrain} The local terrain is rocky and very uneven, especially
along the EW arm with height differences of up to 35~m.  To minimize
the problems of non-coplanarity, it was decided to level the EW arm in
multiples of 64~m (32$\lambda$) so that the antennas in each group
will be at the same height. The height profile of the groups is shown
in the Figure~\ref{fig:fig_2}. For optimal use of the 512-channel
complex correlation receiver, several schemes were considered before
arriving at a configuration consisting of 16 movable trolleys in the
NS arm and 32 groups in the EW arm. Details of these considerations
are given by K. Golap \cite{Golap}.

\paragraph{Interference} When this telescope was conceived, man made
interference in Mauritius was very low. By the time the array became
operational, the increased use of many local communication networks,
had increased man-made interference. The front end of the receiver
system has been built with sufficient bandwidth so that the observing
frequency can be shifted (within 145-155~MHz) to an interference free
zone by tuning the LO. The 1~MHz band around 151.5~MHz has been found
to be relatively quiet and is therefore presently being used.

\subsection {The MRT Array}

\begin{figure}[h]
\epsfig{file=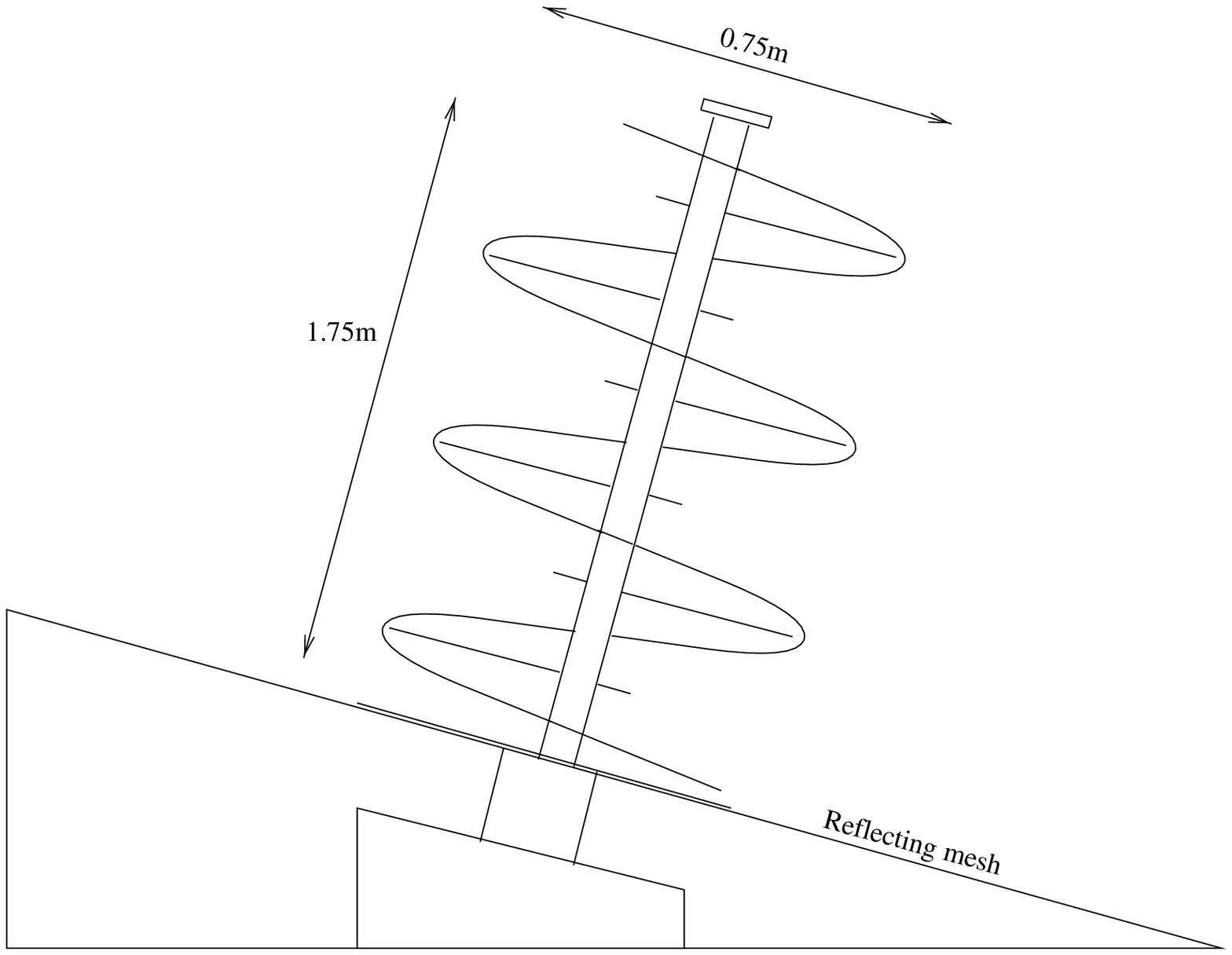,height=10cm,width=14cm} \caption{{\footnotesize
The MRT helix is a peripherally fed monofilar axial mode helix of 3
turns. The helices are mounted with a tilt of $20^{\circ}$ to allow a
better coverage of the southern sky.}} 
\label{fig:fig_3} 
\end{figure}

\paragraph{Helix}The primary element is a peripherally fed monofilar 
\footnote{Monofilar : A term used to distinguish single conductor helix
from helices with two or more conductors} axial-mode helix of 3 turns
with a diameter of 0.75 m and a height of 1.75 m
(Figure~\ref{fig:fig_3}). This is mounted above a stainless steel
reflector mesh of grid size $2''\times 2''$. The helix is wound using
a 1.5 cm diameter round aluminum tubing supported with a central UV
(Ultra Violet) stabilized, light-weight PVC (Poly Vinyl Chloride)
cylinder and radial PVC rods. The axial mode provides maximum
radiation along the helix axis.

The helix responds to frequencies between 100 to 200~MHz with right
circular polarization (IEEE definition). A quarter-wave transformer
used in the feed network optimizes the VSWR to $\leq$1.5 around
150~MHz. The helical antenna, with its reflector, has a canonical
collecting area of about $\lambda^2$ (4 m$^2$ at 150~MHz) with a HPBW
of about $60^{\circ} \times 60^{\circ}$.  The helices are mounted with
a tilt of 20$^{\circ}$ towards the South to get a better coverage of
the Southern sky ($-70^{\circ}$ to $-10^{\circ}$ dec) including the
Southern-most part of the Galactic plane, a region largely unexplored
at meter wavelengths.
        
\paragraph{EW and S arms} The EW arm consists of 1024 helices mounted on a 2~m wide ground plane with an inter-element spacing of 2~m ($\lambda$
at 150~MHz) and is divided into 32 groups of 32 helices each.As
already mentioned, due to the uneven terrain all the EW groups are not
at the same height.  The HPBW of the primary beam of each EW group is
$1.8^{\circ} \times 60^{\circ}$ and allows observation of a source for
roughly $7\times \sec(\delta)$ minutes.  In each group, four helices
are combined using power combiners and the combined output is pre-amplified in
a low noise amplifier (noise temperature $\approx 300$K).  Eight such
amplified outputs are further combined and amplified to form a group
output in the EW array.

\begin{figure}
\epsfig{file=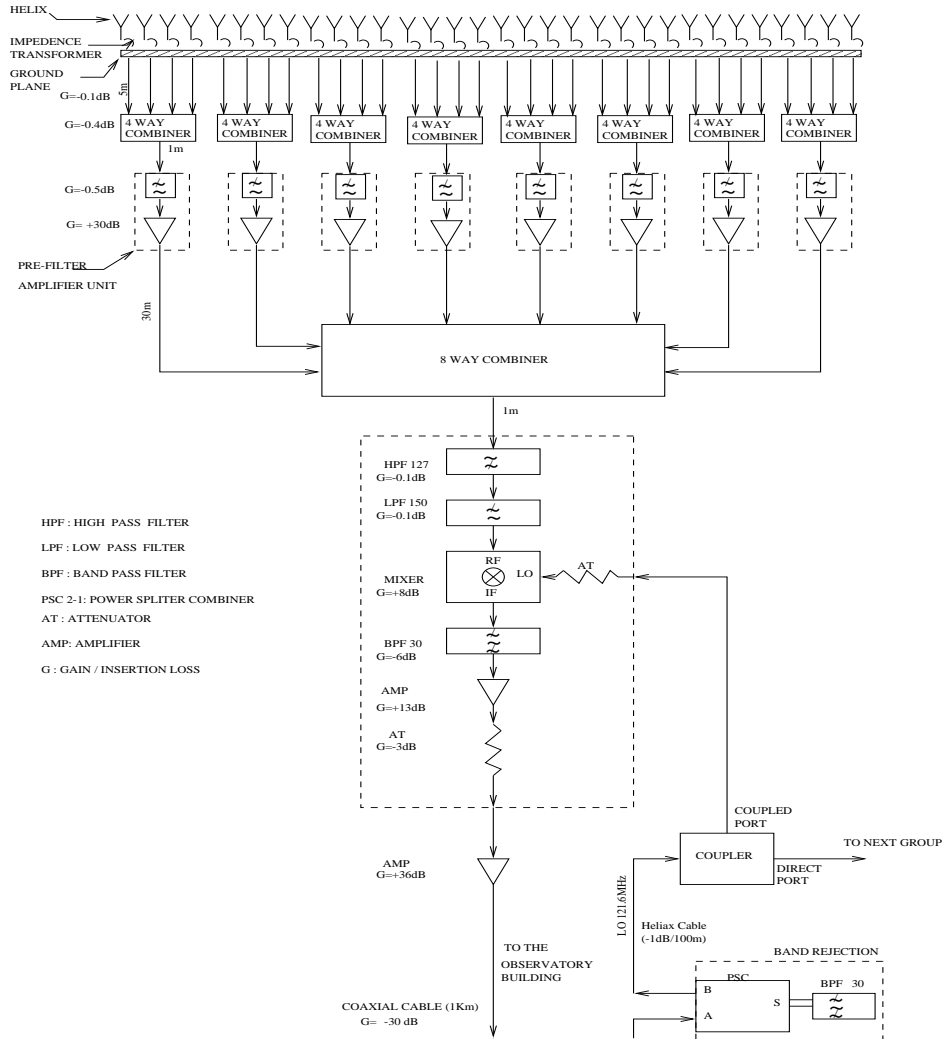,height=14cm,width=12.5cm}
\caption{{\footnotesize Signal path of an EW group. The NS trolley has
a similar flow path except that there are only 4 helices and 1
pre-filter unit}}
\label{fig:fig_4}
\end{figure}

Four helices mounted on a trolley with a 4 m wide ground plane
constitute a group in the S array with a primary beam of HPBW
$15^{\circ}\times 60^{\circ}$. Each of the 32~EW and 16~S group
outputs are heterodyned to an IF of 30~MHz, amplified and transmitted,
using coaxial cables, to the observatory building situated close to
the center of the array. Equal lengths of coaxial cables are used
irrespective of the distance of the groups from the observatory to
ensure that the interferometer outputs are not affected much by
changes in the ambient temperature. The configuration of the EW group
is shown in Figure~\ref{fig:fig_4}. Each NS trolley has a similar
signal flow path except that there are only 4~helices and 1~pre-filter
unit.

\paragraph{Local Oscillator System} Low loss (2.4~dB /100~m at 120~MHz) 
heliax cables running parallel to the East, West and South arms of the
array are used to distribute the LO signal required for
heterodyning. The 121.6~MHz LO signal at each group is obtained by
using directional couplers. To maintain a more or less constant power
level for the LO signal ($-5$~dBm to $-7$~dBm) across the array,
broad-band high-power amplifiers are placed at several places along
the heliax cable.

To minimize the spurious correlations in the visibility measurements,
the following hardware features are built into the LO system.

\begin{itemize}

\item The first LO is phase switched to minimize the effects of cross
coupling. The Clark Lake correlator has the facility for generating
the Walsh functions required for switching. But it is not practical to
use many independent switching waveforms in a series fed LO system
owing to the problem of distribution. Hence, we use a simple
arrangement in which the LO to the first eight southern groups remain
unswitched while, the eastern, the western and the remaining southern
groups are switched with orthogonal square waves. The disadvantage of
this scheme is that the products of the outputs from groups having
same switching signal are prone to spurious correlations (eg. the
product of any two eastern arm group outputs).

\item The signal generator used to produce the LO also generates
spurious signals at other frequencies. The spurious signals around
30~MHz, which is our First IF frequency, leaks to the IF port from the
LO port of the mixer and results in correlation several times the
detection limit. This is minimized by using a band rejection device
centered around 30~MHz in the LO path. Band rejection is achieved
using a band-pass filter centered around 30~MHz and a power-splitter
(Figure~\ref{fig:fig_4}). The filter absorbs the signal around 30~MHz
and reflects the LO required for heterodyning.

\end{itemize}

\subsection {The Receiver System} 

\subsubsection{Second IF and correlator modules} In the 
observatory building the 48 group outputs are further amplified.  Each
of the 32 EW group outputs are split into two using power
dividers. One set of the outputs are combined to form a fan beam of
$4'\times 60^{\circ}$ using phase-shifter modules. This single beam
forming receiver provides a two degree tracking system. Details of
Pulsar observations carried out using this system are described by
N. Issur \cite{nalini}.

\begin{figure}
\epsfig{file=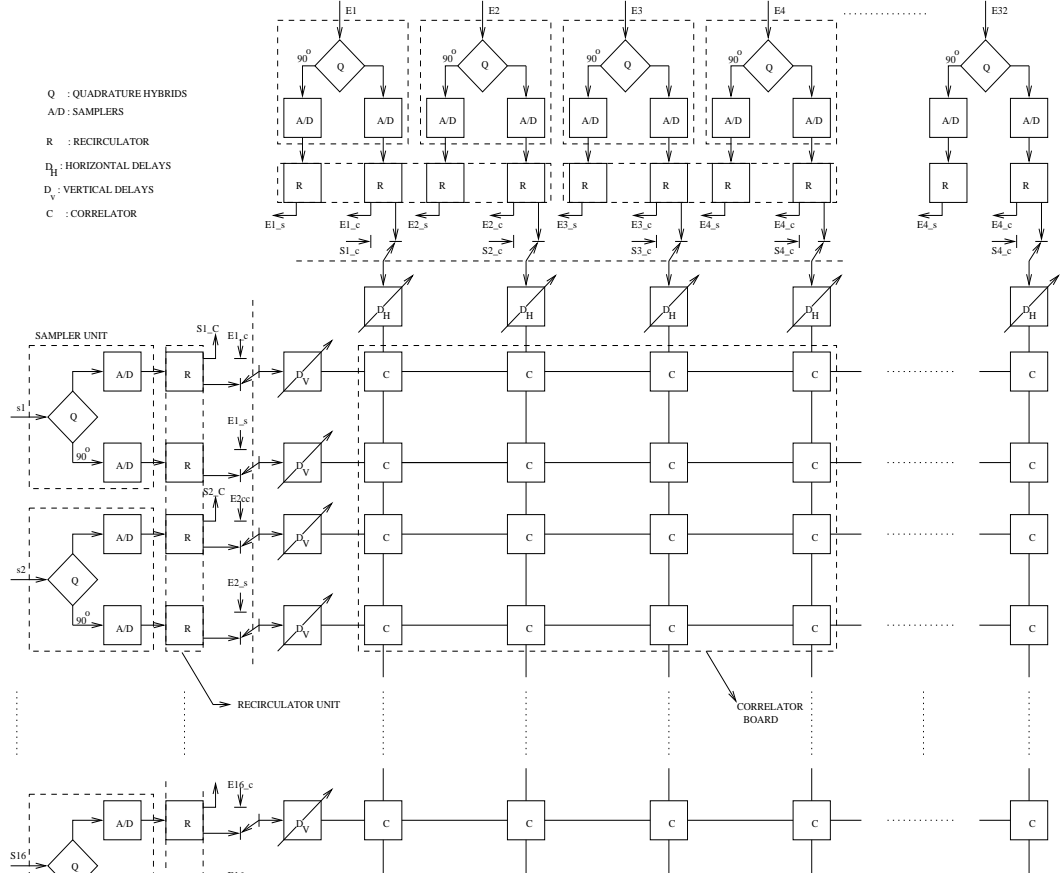,height=14cm,width=12.5cm,angle=270}
\caption{{\footnotesize Diagram showing the general layout of the  signals to
the correlators}}
\label{fig:corr_schem}
\end{figure}

The second set of 32 EW group outputs and the 16 S outputs are down
converted to 10.1~MHz (second IF). Four IF bandwidths, ranging from
0.15 to 3~MHz are selectable. Signals are then fed to an Automatic
Gain Control (AGC) module which keeps the output power level constant.
Each of the 16 group outputs of the S arm and the 16 group outputs of
the E arm is split in a quadrature hybrid to obtain the in-phase and
quadrature components. Then the signals are digitized to 2-bit
3-level, sampled at 12~MHz and fed into the 32 $\times$ 16 complex
correlator to obtain the EW $\times$ S outputs. The high sampling rate
is to reduce the loss of sensitivity due to quantization. More details
of the system are given by Erickson {\it et
al.}\cite{CLRO}. Figure~\ref{fig:corr_schem} shows the general layout
of the samplers, delay lines and correlators.

The last group of the East array (E16) is fed to the correlator in the
place of the $16^{\rm th}$ trolley. This gives a set of baselines
formed between E16 and the E-W array on all observing days. This helps
to check the repeatability of data. These baselines with each trolley
give 31 independent closure information which are used in the
calibration. This mode of observing reduces the number of usable
trolleys to only 15.

Many observing programs require the formation of fan beams
\cite{nalini} corresponding to each arm of the array. The correlator
can also be configured to obtain E $\times$ EW and S $\times$ S
correlations.
	
\paragraph{Self-correlators} Sixty four self-correlators are available
in the receiver system. This provides the necessary information for
obtaining the normalized correlation coefficient from the measured
digital correlation counts using the correction given by D'Addario
{\it et al.}\cite{addario}.

\begin{equation}
\rho= \gamma - \frac{\gamma^3(\alpha_1^2-1)(\alpha_2^2-1)}{6}
\end{equation}
where $ \gamma = (\pi/2)(N/N_{max})\exp{(\alpha_1^2+\alpha_2^2)/2}$.
$ \alpha_1 $ and $ \alpha_2 $ are the $ V_{th}/\sigma $ of the two
channels being correlated and $N/N_{max}$ is the ratio of the
correlation value to the maximum value possible out of the 2-bit
3-level system ($V_{th}$ is the threshold voltage for digitization and
$\sigma$ is RMS value of the signal fed to the digitizer).

The AGCs maintain a constant power level to the samplers, even though
the brightness distribution of the sky changes. Therefore we do not
get the amplitude information of the signal from the sky. This results
in identical correlation for a weak source in a weak background and a
stronger source in a correspondingly stronger background. To measure
the absolute brightness while using the AGC there is a need to measure
the gains.  In many telescopes this is done by adding a fixed amount
of a known signal at each antenna.  At the MRT, however, the variation
in the background radiation as seen by the EW and the S
groups are measured separately by switching off the AGCs, one in each
of the EW and the S arms and using the self-correlators to measure the
total power output of these groups.

The self correlators of the MRT are wired such that they measure the
probability ($P$) that the input signal amplitude, $V$, is between the
threshold levels used for digitization. This probability for a
zero-mean Gaussian signal with RMS fluctuation of $\sigma$ and a
symmetric 2-level digitizer with voltage threshold levels $\pm
V_{th}$, is given by

\begin{equation}
P=\frac{1}{\sigma \sqrt{2\pi}}\int_{-V_{th}}^{+V_{th}} e^{-(\frac{V}{\sqrt{2}\sigma})^{2}} dV =erf(\frac{V_{th}}{\sqrt{2}\sigma}) 
\end{equation}
\noindent knowing $P$, the RMS fluctuation $\sigma$ of the signal can
be obtained. The analog correlation $\rho_a$ is obtained using the
relation

\begin{equation}
\rho_a= \rho\times \sigma_1 \times \sigma_2
\end{equation}

\noindent where $\sigma_1$ and $\sigma_2$ are the RMS of the signals correlated.

In a 2-bit 3-level correlator, sampling the digitized signal at Nyquist
rate, the maximum sensitivity obtainable relative to an analog
correlator is 0.81. This is obtained when $V_{th}/\sigma $ is
0.61. The sensitivity changes by only 5\% from this optimal value when
the signal power changes by 40\% \cite{bowers}. Hence, switching off
the AGC does not affect the signal-to-noise ratio (SNR) of the
channels used to measure the total power as the variation of the sky
brightness is less than $40\%$.

\paragraph{Recirculator} Although the use of larger bandwidths results
in better sensitivity of a telescope, it restricts the angular range
over which an image can be made if the relative delays between the
signals being correlated are not compensated. When the uncompensated
delay between the signals becomes comparable to the inverse of the
bandwidth used, the signals will be decorrelated. At MRT we normally
use a bandwidth of 1~MHz.  Since the EW group has a narrow primary
beam of two degrees in RA, this bandwidth does not pose a problem for
synthesizing the primary beam in this direction. However both the EW
and NS groups have wide primary beams in declination extending from
$-70^{\circ}$ to $-10^{\circ}$. For signals arriving from zenith
angles greater than 10 degrees on NS baselines longer than 175 m, the
uncompensated delay results in bandwidth decorrelation greater than
20\%.


To overcome this loss of signal one has to measure visibilities with
appropriate delay settings, while imaging different declination
regions.  To keep the loss of signal for a bandwidth of 1~MHz to less
than 15\% in the entire declination range $-70^{\circ}$ to
$-10^{\circ}$, the longer baselines have to be measured with four
delay settings.

With the existing correlator system this implied observations for four
days at trolley locations measuring long baselines. Furthermore, to
be able to make interference free maps, observations have to be
repeated approximately 3 times at each location. This would make the
time required to complete the survey very large.


To solve this problem, a recirculator system which measures
visibilities with different delay settings using the available
correlators in a time multiplexed mode has been incorporated in the
receiver system\cite{sachdev}.  In this system, the data is sampled at
3~MHz, stored in a buffer memory and the correlations are measured at
12~MHz rate. Such processing by the correlator at a higher speed than
the input sampling rate allows correlations to be measured with four
delay settings.  This ensures that the observations at each trolley
allocation, even at longer baselines, can be carried out in one day,
thereby, improving the surveying sensitivity. The loss of sensitivity
due to the decrease in the input sampling rate from 12~MHz to 3~MHz is
not serious. The recirculator system is described in detail by
S. Sachdev \cite{sachdev}.

The recirculator system can configure the correlator in the $
(E+W)\times E $ and $S \times S$ modes for one integration time
(roughly 100 ms in every second of data collection). Since both EW and S
arrays are linear, this mode provides visibilities on several
redundant baselines to calibrate the array using the Redundant
Baseline Calibration technique \cite{noordam}. Presently the software
at MRT does not make use of this data for array calibration.

\subsection{Observations} 

In the T configuration, we are interested in sampling baselines with
north-south components from 0~m to 880~m with a sampling of 1~m
(except the 1~m spacing~[Section~\ref{sub_sec-design_crit}]). This
ensures that the grating response will be confined well outside the
primary beam response. To measure the visibilities with a 1~m spacing
using 15 trolleys however requires 60~days of observing. Each day the
trolleys are spread over 84~m with an inter-trolley spacing of
6~m. 512 complex visibilities are recorded with 1~sec of integration.
After obtaining data for 24~sidereal-hours the trolleys are moved by
one meter. After 6~days of successful observing we get visibilities
measured over 90~m. The trolleys are then moved en~bloc to measure the
next set of visibilities.

\section{Imaging with the MRT}

\begin{figure}
\epsfig{file=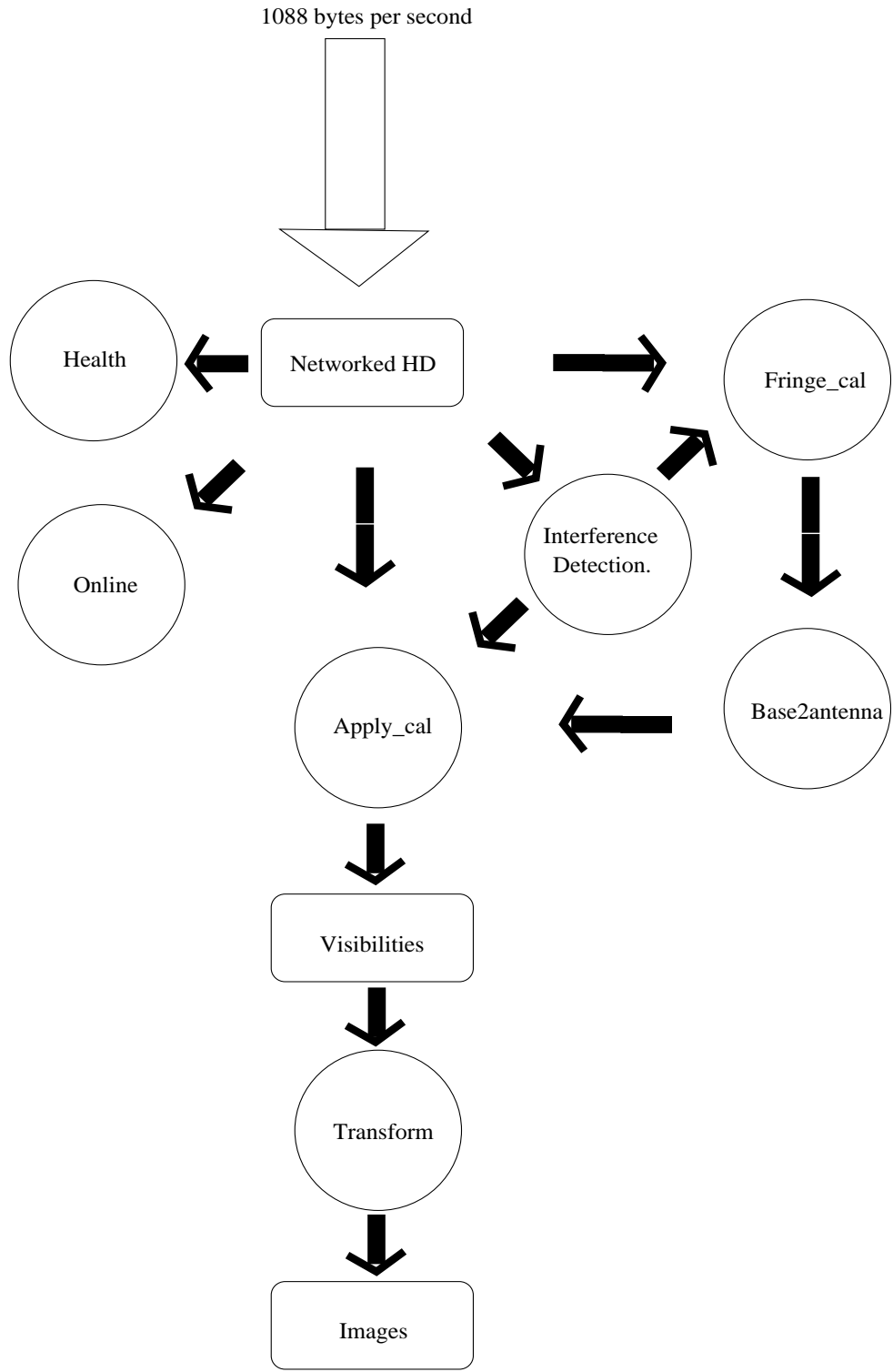,height=20cm,width=13cm}
\caption{\footnotesize{Flow Diagram of data acquisition, checking and
processing.}}
\label{fig:flow}
\end{figure}

The visibilities measured are processed off-line using
MARMOSAT\footnote{A marmoset is a small monkey, distantly related to
the ape.}, the MAuRitius Minimum Operating System for Array
Telescopes\cite{Dodson}. This is designed in-house to transfer the
visibilities to images which can be ported to AIPS. The flow diagram
showing the various processing stages is shown in
Figure~\ref{fig:flow}.


The programs in MARMOSAT cater to the following needs: detection of
bad data, determination of complex gain of the antennae, position
calibration of antennas, combining data of different days,
transforming the visibilities to get brightness distribution.

The program {\bf Online} presents an uncalibrated view of the data. It
is designed to read visibilities from several baselines and offers to
display many functions such as phase, magnitude, closure and fringe
frequencies of the visibility measured. The program {\bf Health}
monitors the signal-to-noise ratio of the visibilities, correlator
offsets, magnitudes of the visibility of all the baselines and
displays the mean and the scatter. This helps to detect gross errors
quickly. The program {\bf Interference Detection} detects the
interference points which are rejected in subsequent processing. The
details of interference detection and their statistics at MRT are
given by S. Sachdev\cite{sachdev}. The programs {\bf Fringe-cal} and
the {\bf Transform} are very special to the MRT due to its
non-coplanarity and configuration. The following sections describe the
concepts behind these programs. Further details of these programs are
given by R. Dodson\cite{Dodson}.

\subsection{Antenna calibration} The complex gains of
the 47 antennas\footnote {Since the last group of the Eastern array
(E16) is used as an input to the correlator in the place of the
sixteenth trolley.} are estimated using the measured visibilities for
the calibrators MRC0915-119, MRC1932-464 and MRC2211-173. The
sensitivity per baseline is 30~Jy for an integration time of one
second.  In the $1.8^{\circ}$ EW-group primary beam, the above sources
are observable in each baseline with a minimum SNR of about 35.

One of the advantages of observing in the Southern sky is that the two
very strong sources, Cyg~A and Cas~A are not in the primary beam, where
they would dominate the system temperature causing dynamic range
problems.





The disadvantage, however, is a paucity of strong sources required for
reliable calibration. This leaves us with a situation where we can
calibrate the array only a few times in a sidereal day.

The visibility phase is estimated by fringe fitting (hence the name
{\bf Fringe-cal} for the program), where we assume that the sky in the
primary beam is dominated by the calibrator. The instrumental phase
estimated is simply the difference between the phase of the observed
visibilities and the expected geometric phase due to the point source
calibrator (or calibrators). The instrumental gain is estimated by
measuring the relative amplitudes of the fringes on different
baselines. The instrumental phase is insensitive to short term
interference and also to fringes due to other sources in the sky,
provided their fringe rate is significantly different from that of the
calibrator for a given baseline. Therefore, at short baselines where
there are less than two fringes in the EW-group beam due to the
calibrator, satisfactory calibration cannot be obtained. 

This is tackled at MRT by calculating 47 antenna complex gains from
the measured 512 complex visibilities ({\bf base2antenna} in
Figure~\ref{fig:flow}). At this stage we reject those baselines with
short EW components (essentially the first four E and W group outputs
with the S array) and input the closure information obtained from
multiplying the E16 group with the EW and S array group outputs.

We have found that the day-to-day RMS variation per baseline in phase
is about $ \pm 7^{\circ}$ and the RMS amplitude variation is about $
\pm 0.1$ dB.  The RMS variation of phase from one calibrator to
another (eg. MRC1932-464 and MRC0915-119) is about $\pm 10^{\circ}$.

\paragraph{Transforming Visibilities to Brightness}

The {\em u,v} coverage of MRT (EW and the NS components of the
baseline) can be thought of as a pleated sheet, extended in both {\em
u} and {\em v}, with discrete steps in {\em w}~(height) as we move
from one EW group to the one at a different height. As we are imaging
a very large field of view ($60^{\circ}$ field), the approximate
coplanar approach, wherein the phase term due to the heights is
assumed to be a constant over the synthesized field of view, is
invalid. Thus for MRT, a 3D imaging method is required.

Here we transform the visibilities using a Fast Fourier Transform
(FFT) along the regularly sampled {\em v} axis, apply a Direct Fourier
Transform (DFT) along the {\em w} axis and finally sum along {\em u}
to obtain the image on the meridian. A DFT on {\em w} is required as
the sampling is not uniform. The direct transform corrects
every term along the zenith angle on the meridian for the group
heights. This is equivalent to phasing the groups to a common (and
artificial) 2D plane.

The NS array slopes downwards at about $ 1/2 ^{\circ}$ to the
horizontal till about 655~m, and then slopes upwards at about $1
^{\circ}$ to the horizontal. These two parts are treated separately
while transforming. The slope is taken into account by assuming an
instrumental zenith appropriate to the slope of the track. While
combining contributions from different parts we introduce necessary
corrections to get an instrumental zenith equal to the latitude of the
place.


The sampling of the visibilities on the EW grid is at intervals equal
to the size of each EW group. This gives a grating response which falls
on the nulls of the primary beam while synthesizing on the
meridian. For synthesis away from the meridian, one of the grating
response starts moving into the primary beam. This leads to the
synthesized beam being  a function of the hour angle. Hence to simplify
matters, imaging is presently done on the meridian only.

\subsection{Initial Results} One set of observations for all the 880
trolley positions have been made. Because of the presence of the Sun
and the day time interference, the data set is not complete for the
full RA coverage. A second set of data is being taken to cover the
complete accessible sky. The observations are expected to be completed
in 1998.

\begin{figure}
\epsfig{file=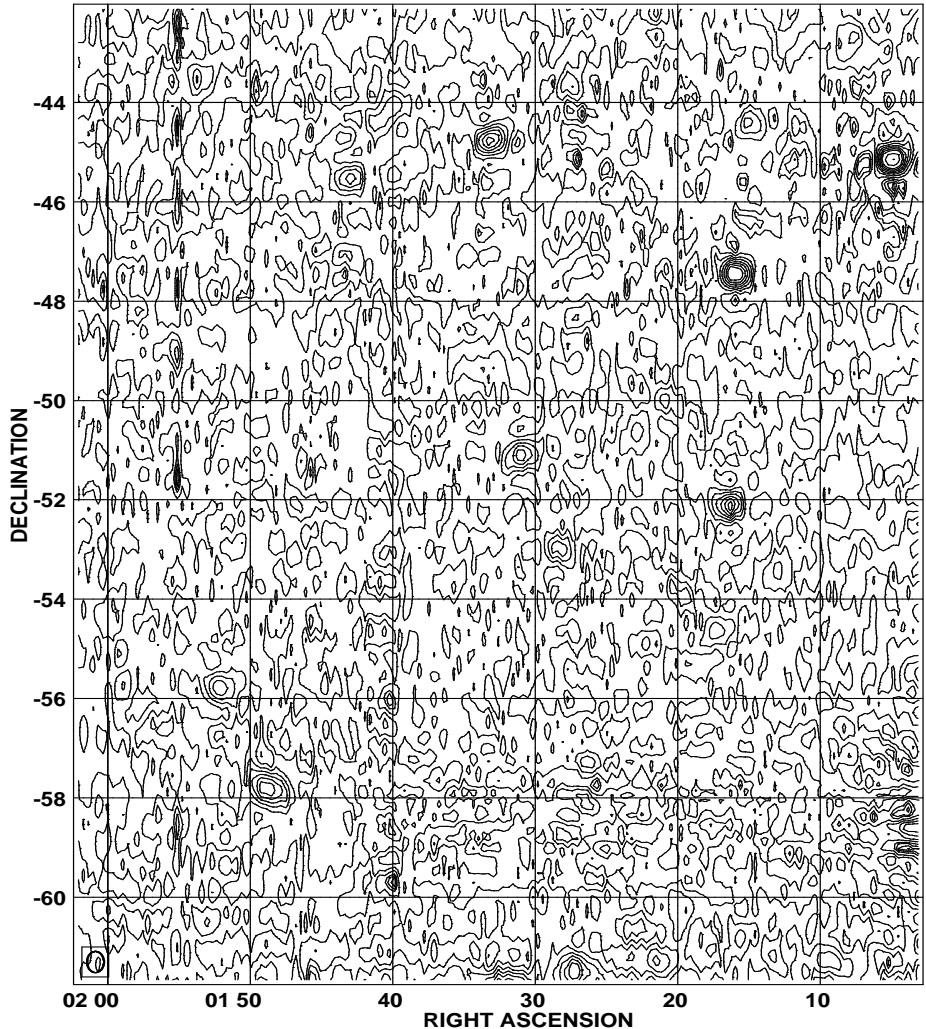,height=14cm,width=12.5cm}
\caption{\footnotesize {Map at 13' x 18' resolution of a typical region away from the galactic plane. The contours are -9 -7 -5 -4 -3 -2 -1 1 2 3 4 5 7 9 11 13 15 17 19 23 27 31 35 45 55 65~Jy/beam. The coordinates are for epoch 2000}}
\label{fig:FIG3}
\end{figure}

\begin{table}
\begin{center}
\begin{tabular}{|c|c|c|}
\hline
\multicolumn{1}{|c|}{RA} & \multicolumn{1}{|c|}{Declination} & \multicolumn{1}{|c|}{ Flux Density (Jy)} \\
\hline
 01:05:26 & -45:04 &  20 \\ \hline
 01:07:51 & -42:40 &  6 \\ \hline
 01:08:53 & -52:20 &  4 \\ \hline
 01:13:23 & -53:23 &  3 \\ \hline
 01:15:29 & -44:23 &  4 \\ \hline
 01:16:30 & -47:24 &  15  \\ \hline
 01:16:51 & -52:08 &  10  \\ \hline
 01:17:53 & -54:39 &  3  \\ \hline
 01:18:54 & -62:42 &  7  \\ \hline
 01:19:58 & -53:48 &  3  \\ \hline
 01:21:22 & -49:59 &   4  \\ \hline
 01:23:27 & -50:42 &    5 \\ \hline
 01:27:15 & -51:37 &   4  \\ \hline
 01:27:34 & -61:32 &   9  \\ \hline
 01:27:57 & -48:17 &   4  \\ \hline 
 01:28:38 & -52:58 &   5  \\ \hline
 01:31:24 & -51:06 &   8  \\ \hline
 01:31:44 & -61:53 &   12  \\ \hline
 01:33:30 & -44:40 &  10  \\ \hline
 01:43:13 & -45:27 &  6  \\ \hline
 01:49:04 & -57:47 &    8  \\ \hline
 01:52:33 & -55:49 &   4  \\ \hline
 01:54:19 & -43:32 &   5  \\ \hline
 01:55:41 & -49:05 &   4 \\ \hline
 01:59:30 & -47:30 &   5 \\ \hline
\end{tabular}

\caption{Table giving point sources detected in the image of Figure~\ref{fig:FIG3},
the coordinates are for epoch 2000}
\label{tab:tab_source}
\end{center}
\end{table}

For imaging on the meridian with all the 880~m baselines, the point
source sensitivity will be about 200~mJy(3 $\sigma$) at the peak of
primary beam with 8~seconds of integration and a bandwidth of
1~MHz. The confusion limit for an instrument of resolution $ 4' \times
4'$ is about 13~mJy. For extended structures the surface brightness
sensitivity is about~5~mJy/arcmin$^2$.

While the data collection is under way, we are processing the
available data to make a low resolution survey of the
sky. Observations up to a baseline of 178~m do not require the
recirculator. We have taken this as a natural cutoff and have made low
resolution images using only 8 central groups of the EW arm and
trolley positions up to 178~m in the NS. A part of the survey covering
the RA range 18:00 to 24:00 hrs and 00:00 to 05:00 hrs has been
completed. In this paper, we present a deconvolved image, of the
region extending from 0100 hrs to 0200 hrs in RA and $-62^{\circ}$ and
$-42^{\circ}$ in declination, with a resolution of $13' \times 18'
\sec(\delta + 20.14^{\circ})$. MRC1932-464, with a flux density of
81~Jy/beam was used as the primary flux calibrator \cite{Golap}. The
expected RMS noise due to confusion and system temperature (with an
integration time of 19 seconds) are expected to be 0.7~Jy. The noise
seen on the map is 0.8~Jy and is very close to the expected
value. More than 25 point sources in the flux density range 3 to 20~Jy
can be seen in the map. Many of them have been identified with
Molongolo and Culgoora sources. A comparison of the flux densities of
the sources common to MRT and Culgoora lists showed that the error on
the estimated flux densities is less than 25\%. The vertical line at
RA~01h55m and the feature at RA~01h02m, Dec~$-59^{\circ}$ are due to
man made interference.  The image has been presented here for
completeness and to illustrate the success of the methods developed
for imaging. A separate paper on the low resolution survey will
discuss more details such as the deconvolution procedure for a
non-coplanar array like MRT, method used for detecting point sources,
measurement of the zero spacing and the flux density scale used.

\section{Acknowledgment}

We thank V. Radhakrishnan for his interest and encouragement from the
initial stages of the project. Our thanks are due to W. E. Erickson
for providing the CLRO correlator to MRT. We gratefully acknowledge
M. Modgekar, C.M. Ateequlla and H.A. Aswathapa from the Raman Research
Institute, R. Somanah and N. Issur from the University of Mauritius
and G.N. Rajashekar from the Indian Institute of Astrophysics for
their efforts in this project. We thank all those from the Raman
Research Institute, Indian Institute of Astrophysics, and the
University of Mauritius who have contributed towards this
project. ChVS thanks J. E. Baldwin for many useful discussions and
suggestions.

 


\bibliography{paper}

\begin{thebibliography}{10}

\bibitem{6C_paper1}
J.E. Baldwin, R.C. Boysen, J.E. Hales, S.E.G.~Jennings, P.C. Waggett, P.J.
  Warner, and D.M.A. Wilson.
\newblock The 6c survey of radio sources -- 1.
\newblock {\em Mon. Not. R. astr. Soc.}, 217:717--730, 1985.

\bibitem{bowers}
F.K. Bowers and R.J. Klinger.
\newblock Quantization noise of correlation spectrometers.
\newblock {\em Astr. Astrophys. Suppl.}, 15:373--380, 1974.

\bibitem{addario}
L.R. D'Addario, A.R. Thompson, F.R. Schwab, and J.~Granlund.
\newblock Complex cross correlators with 3-level quantization - design
  tolerances.
\newblock {\em Radio Science}, 19(3):931--945, 1984.

\bibitem{Dodson}
R.~Dodson.
\newblock {\em The Mauritius Radio Telescope and a Study of Selected Super Nova
  Remnants Associated with Pulsars}.
\newblock PhD thesis, University of Durham, 1997.

\bibitem{CLRO}
W.C. Erickson, M.~J. Mahoney, and K.~Erb.
\newblock The clark lake teepee-tee telescope.
\newblock {\em Astrophys. J. Suppl.}, 50(403-419), 1982.

\bibitem{Golap}
K.~Golap.
\newblock {\em Synthesis Imaging at 151.5 MHz using the MRT}.
\newblock PhD thesis, University of Mauritius, 1998.

\bibitem{PMN}
M.R. Griffith and A.E. Wright.
\newblock {\em Astron. J.}, 105:1666, 1993.

\bibitem{nalini}
N.~Issur.
\newblock Pulsar observation with mrt.
\newblock {\em Bull. Astr. Soc. India}, 1997.

\bibitem{Mills}
B.Y. Mills, O.B. Slee, and E.R. Hill.
\newblock A catalogue of radio sources.
\newblock {\em Aust. J. Phys.}, 11:360, 1958.

\bibitem{noordam}
J.~E. Noordam and A.G. de~Bruyn.
\newblock High dynamic range mapping of strong radio sources,with application
  to 3c84.
\newblock {\em Nature}, 299:597, 1982.

\bibitem{reber}
G.~Reber.
\newblock {\em Astrophys. J.}, 100:279, 1944.

\bibitem{sachdev}
S.~Sachdev.
\newblock {\em Wide Field imaging with the Mauritius Radio Telescope(under
  preparation)}.
\newblock PhD thesis, University of Mauritius, 1998.

\bibitem{culgoora}
O.B. Slee.
\newblock Culgoora-3 list of radio source measurements.
\newblock {\em Aust. J. Phys. Astrophys. Suppl.}, 43:1--123, 1977.

\end{thebibliography}
\bibliographystyle{plain}
\end {document}